\documentclass[apj]{emulateapj}
\usepackage{apjfonts}

\usepackage{epsfig}
\usepackage{natbib}          
\shorttitle{43 and 86 GHz polarimetry of 3C273}
\shortauthors{Attridge et al}

\begin{document}

 \title{CONCURRENT 43 AND 86 GHz VERY LONG BASELINE POLARIMETRY OF 3C273}

\author{Joanne M.~Attridge\altaffilmark{1}}
\author{John F.~C.~Wardle\altaffilmark{2}}
\author{Daniel C.~Homan\altaffilmark{3}}

\altaffiltext{1}{MIT Haystack Observatory, Off Route 40, Westford, MA 01886;
jattridge@haystack.mit.edu}
\altaffiltext{2}{Physics Department MS057, Brandeis University, 
Waltham, MA 02454; wardle@brandeis.edu}
\altaffiltext{3}{Jansky Fellow, National Radio Astronomy Observatory; present address Department of Physics and Astronomy, Denison University, 
Granville, OH 43023; homand@denison.edu}

\begin{abstract}

We present sub-milliarcsecond resolution total intensity and linear 
polarization VLBI images of 3C\,273, using concurrent 43 and 86~GHz data 
taken with the Very Long Baseline Array in May 2002. The structure seen in the innermost jet suggest that we have fortuitously caught the jet in the act of changing direction.
The polarization images confirm that the core is unpolarized (fractional 
polarization $\le 1\%$) at 86 GHz, but also show well ordered magnetic 
fields ($m \sim 15\%$) in the inner jet, at a projected distance 
of 2.3 pc from the core. In this strongly polarized region, the rotation measure 
changes across the jet 
by $\sim 4.2 \times 10^{4}$~rad m$^{-2}$ over an angular width of 
about $0.3$ milliarcseconds. If the lack of polarization in the core is 
also attributed to a Faraday screen, then a rotation measure dispersion
$\gtrsim 5.2 \times 10^{4}$~rad m$^{-2}$ must be present in or in front 
of that region. These are among the highest rotation measures reported so far in the nucleus of any active galaxy or quasar, and must occur outside (but probably close to) the radio emitting region. The transverse rotation measure gradient is in the same sense as that observed by Asada et al and by Zavala and Taylor at greater core distances. The magnitude of the transverse gradient decreases rapidly with distance down the jet, and appears to be variable.

\end{abstract}



\section{Introduction}
Multi-frequency Very Long Baseline Polarimetry (VLBP) observations are
crucial for determining the magnetic field structure and the 
distribution of Faraday rotating material in the innermost few parsecs 
of the cores of active galaxies and quasars.

The first successful VLBP observations at 86~GHz were reported by \citet{att01}, who presented polarized images of the well-known quasars 3C\,273 ansd 3C\,279. Here we report 
concurrent 43~GHz ($\lambda6.9$mm)  and
86~GHz ($\lambda3.5$mm) VLBP of the 3C\,273 that allow us
to investigate the distribution of Faraday rotating material near
the core of this object, probing linear scales as small as a few light
years\footnote{$H_{0}=70$~km~s$^{-1}$~Mpc$^{-1}$, $\Omega_{M}=0.3$,
and $\Omega_{\Lambda}=0.7$ are assumed in all calculations.}.

3C\,273 (J1229+0203) is a good source on which
to attempt 86~GHz VLBP, as it is one of
the brightest quasars in the sky at that frequency. Also, its low
redshift of $z=0.158$
gives very high linear resolution (2.73~pc/mas) at the source. 
It is also a well studied superluminal source. For instance, Homan et al (2001) measured the proper motions of five jet components over six epochs in 1996 at both 15 and 22~GHz. Results ranged from 0.77 to 1.15~mas~yr$^{-1}$, corresponding to superluminal motions of
$\beta_{app} = 7.9-11.9$.

\section{Observations and Calibration}
Concurrent 43 and 86~GHz observations of 3C\,273 were performed
on 9 May 2002 (epoch 2002.35) using the Very Long Baseline Array
(VLBA)\footnote{The National Radio Astronomy Observatory is a facility
of the National Science Foundation, operated under a cooperative agreement
by Associated Universities, Inc.} over 10.5~hours.
The VLBA was equipped with all ten 43~GHz receivers and seven 86~GHz
receivers (Fort Davis, Kitt Peak, Los Alamos, Mauna Kea, North Liberty,
Owens Valley, Pie Town), effectively yielding the same resolution at each
frequency. With a CLEAN beam of about $0.5 \times  0.25$ mas, the linear resolution at the source is about $6 \times 3$ light years.  

The data were recorded with a total bandwidth of 64~MHz, and
were correlated at the National Radio Astronomy Observatory (NRAO)
in Socorro, NM. Approximately $1.5$ hours total integration
was spent on 3C\,273 at 43~GHz, and $3.0$ hours at 86~GHz. These
integration times lead to theoretical RMS noise values of
$\sim$0.5~mJy beam$^{-1}$ at 43~GHz, and $\sim$2~mJy beam$^{-1}$
at 86~GHz.
The observations of 3C\,273 were interleaved with those of the quasar
3C\,279, which was a second target source and served to confirm our
calibration. Results for 3C\,279 will be presented in a separate
paper.

\begin{figure*}[t]
\figurenum{1}
\plottwo{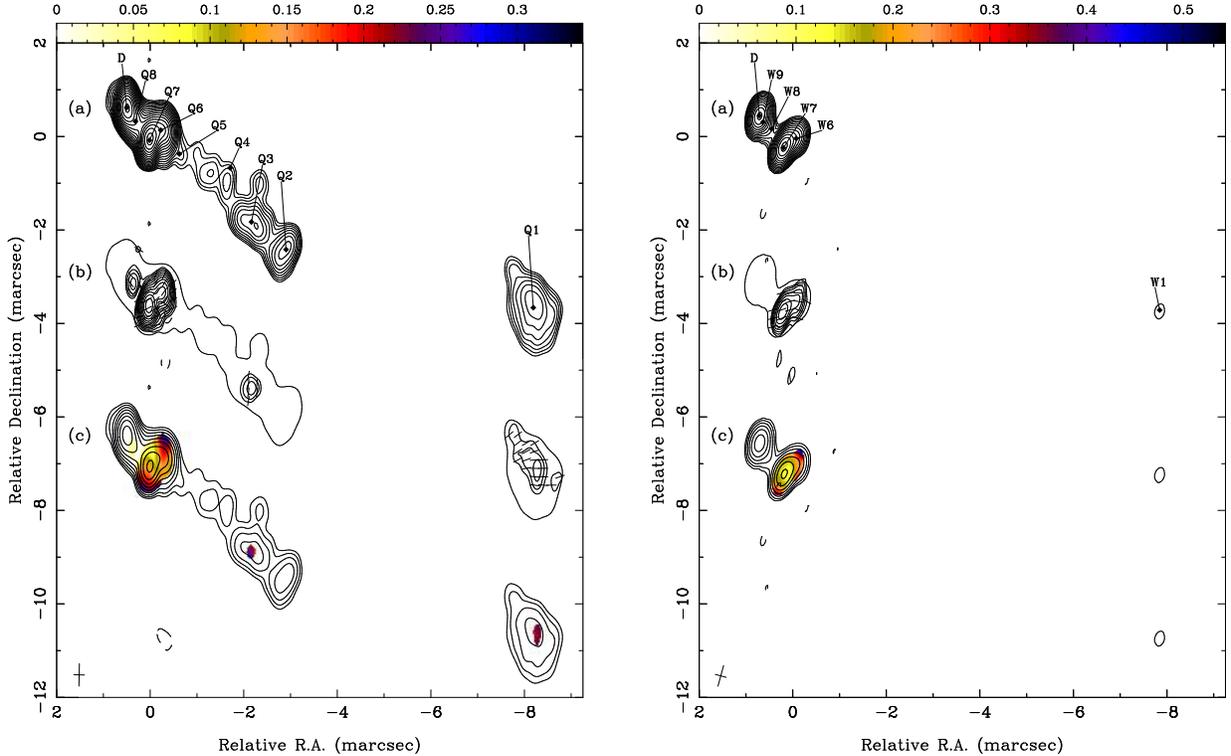}{big_whole_3c273_3mm.ps}
\caption{Naturally weighted images of 3C\,273,
epoch 2002.35, made with the VLBA at 43~GHz (left) and 86~GHz (right). In each panel we display ($a$) Total intensity 
distribution,($b$) Linear polarization distribution, with tick marks
showing the orientation $\chi$ of the electric vectors, and ($c$) Fractional polarization
distribution in grayscale (in color on astro-ph), over every other $I$ contour. The restoring beams are shown in the lower left corners. At 43~GHz, the contours start at 15 mJy/beam and increase by factors of $\sqrt{2}$ up to peak values of 4547 mJy/beam (total intensity) and 630 mJy/beam (polarized intensity). At 86~GHz, the contours start at 25 mJy/beam and increase by factors of $\sqrt{2}$ up to peak values of 1815 mJy/beam (total intensity) and 256 mJy/beam (polarized intensity).
\label{fig1}}
\end{figure*}

The NRAO AIPS and Caltech DIFMAP packages were used for calibration
and hybrid imaging. Details about VLBP calibration and imaging
may be found in \citet{cot93} and Roberts, Wardle, \& Brown (1994). 
Methods to account for the complexities of
86~GHz calibration not present in lower frequency
data, and to check the validity of the 86~GHz solutions are described in
\citet{att01} and \citet{att99b}.

The calibration of the instrumental polarization
(D-terms) was performed for 3C\,273 and 3C\,279
separately, and results were compared. At 43~GHz, the
typical D-term values were found to be $\sim 0.03$, and at 86~GHz they
were $\sim 0.11$. The r.m.s. {\em difference} between the D terms 
determined using 3C\,273 and those determined using 3C\,279 was $0.02$ at 86~GHz. 
This is a measure of the errors in the determination of the D terms. 
Following the prescription in Appendix 1 of Roberts et al (1994), we 
estimate that the corresponding error in fractional 
polarization in the images is $\sim 0.003$ ($0.3\%$). This is consistent with the 
observed fractional polarization of the core component of 3C\,273, which 
is $\le 1.0\%$.

\section{Images}
Naturally weighted images of 3C\,273 in total intensity, polarized intensity, and fractional
linear polarization at 43~GHz and 86~GHz are shown in Figure~1. Tick marks represent
the orientation of the electric vectors of the polarized radiation. The orientations of the EVPAs at 43~GHz were derived using data from the ``VLA/VLBA 
Polarization Calibration Page''
\citep{tay00b}, in which 3C\,273 was observed only one day prior to our own observations.

Currently, there is no source to use as a calibrator to set the zero-point of the EVPAs at 86~GHz. We therefore arbitrarily chose to rotate tick marks in the 86~GHz $P$ image so that the EVPA of component 7 is the same at 43 and
86~GHz. To see the EVPAs more easily, Figure~2 
shows a zoom of the polarized emission from the core region at each frequency.

Model fits to the u-v data are presented
in Table~1; component labels for 86~GHz were matched to corresponding
43~GHz components for ease of comparison.

The 43~GHz total intensity image has a SNR
of 2400, and a peak to lowest contour dynamic range of $\sim$300:1. The
86~GHz $I$ image presented in Figure~2$a$ has a SNR of 590, and a peak
to lowest contour dynamic range of $\sim$75:1. The corresponding numbers 
for the $P$ images are about $7$ times smaller.

\begin{deluxetable}{ccccccccc}
\tabletypesize{\scriptsize}
\tablecaption{Epoch $2002.35$ Modelfit Results for 3C\,273\label{tbl-1}}
\tablewidth{0pt}
\tablehead{
\colhead{} & \colhead{} & \colhead{} & \colhead{} & \colhead{} & \colhead{} &
\colhead{Major} & \colhead{Minor} & \colhead{}\\
\colhead{Com-} & \colhead{$r$} & \colhead{$\theta$} &
\colhead{$I$} & \colhead{$m$} & \colhead{$\chi$} &
\colhead{axis} & \colhead{axis} & \colhead{$\phi$} \\
\colhead{ponent} & \colhead{(mas)} & 
\colhead{(deg)} &
\colhead{(Jy)} & \colhead{(\%)} & \colhead{(deg)} &
\colhead{(mas)} & \colhead{(mas)} & \colhead{(deg)} \\
\colhead{(1)} & \colhead{(2)} & \colhead{(3)} & \colhead{(4)} &
\colhead{(5)} & \colhead{(6)} & \colhead{(7)} &
\colhead{(8)} & \colhead{(9)} 
}
\startdata
\multicolumn{9} {c} {43 GHz} \\ 
\hline
 & & & & & & & & \\
  D & $\ldots$ & $\ldots$ & $3.03$ & $\le 0.2$ & $\ldots$ & $0.1$ & $0.1$ & $40$ \\
 Q8 & $0.35$ & $-148$ & $1.32$ & $6.7$ & $28$ & $0.3$ & $0.1$ & $35$ \\
 Q7 & $0.85$ & $-145$ & $4.60$ & $14.8$ & $-63$ & $0.1$ & $< 0.1$ & $-47$ \\
 Q6 & $0.87$ & $-124$ & $2.95$ & $15.6$ & $-31$ & $0.3$ & $0.1$ & $80$ \\
 Q5 & $1.50$ & $-132$ & $0.08$ & $\ldots$ & $\ldots$ & $< 0.1$ & $< 0.1$ & $\ldots$ \\
 Q4 & $2.56$ & $-120$ & $0.08$ & $\ldots$ & $\ldots$ & $< 0.1$ & $< 0.1$ & $\ldots$ \\
 Q3 & $3.62$ & $-133$ & $0.91$ & $9.5$ & $-28$ & $1.5$ & $0.4$ & $43$ \\
 Q2 & $4.56$ & $-132$ & $0.39$ & $\ldots$ & $\ldots$ & $0.7$ & $0.2$ & $-28$ \\
 Q1 & $9.69$ & $-116$ & $0.90$ & $23.3$ & $-82$ & $0.9$ & $0.6$ & $19$ \\
 & & & & & & & & \\
\hline
\multicolumn{9} {c} {86 GHz} \\ 
\hline
 & & & & & & & & \\
  D & $\ldots$ & $\ldots$ & $1.37$ & $\le 1.0$ & $\ldots$ & $0.1$ & $< 0.1$ & $55$ \\
 W9 & $0.16$ & $-151$ & $0.16$ & $\ldots$ & $\ldots$ & $< 0.1$ & $< 0.1$ & $\ldots$ \\
 W8 & $0.39$ & $-137$ & $0.12$ & $\ldots$ & $\ldots$ & $< 0.1$ & $< 0.1$ & $\ldots$ \\
 W7 & $0.86$ & $-143$ & $2.09$ & $16.7$ & $-63^{*}$ & $0.2$ & $0.1$ & $-35$ \\
 W6 & $0.93$ & $-122$ & $0.71$ & $16.2$ & $-97$ & $0.2$ & $< 0.1$ & $70$ \\
 W3 & $3.53$ & $-135$ & $0.03$ & $\ldots$ & $\ldots$ & $< 0.1$ & $< 0.1$ & $\ldots$ \\
 W2 & $4.62$ & $-131$ & $0.05$ & $\ldots$ & $\ldots$ & $0.5$ & $0.1$ & $-21$ \\
 W1 & $9.52$ & $-116$ & $0.07$ & $\ldots$ & $\ldots$ & $0.5$ & $< 0.1$ & $27$ \\
\enddata
\tablecomments{Columns are as follows:  (1) 
component name;
(2) distance from
easternmost feature D; (3) position angle of separation from D; (4) total
intensity; (5) fractional linear polarization;
(6) orientation of the
linear polarization position angle; (7) major axis of the model component;
(8) minor axis of the model component; (9) orientation of the major axis.\\
$ ^{*}$ All EVPAs at 86 GHz have been arbitrarily rotated so that the 86 GHz and 43 GHz values for this component are aligned.}
\end{deluxetable}

\begin{figure}[b]
\figurenum{2}
\plotone{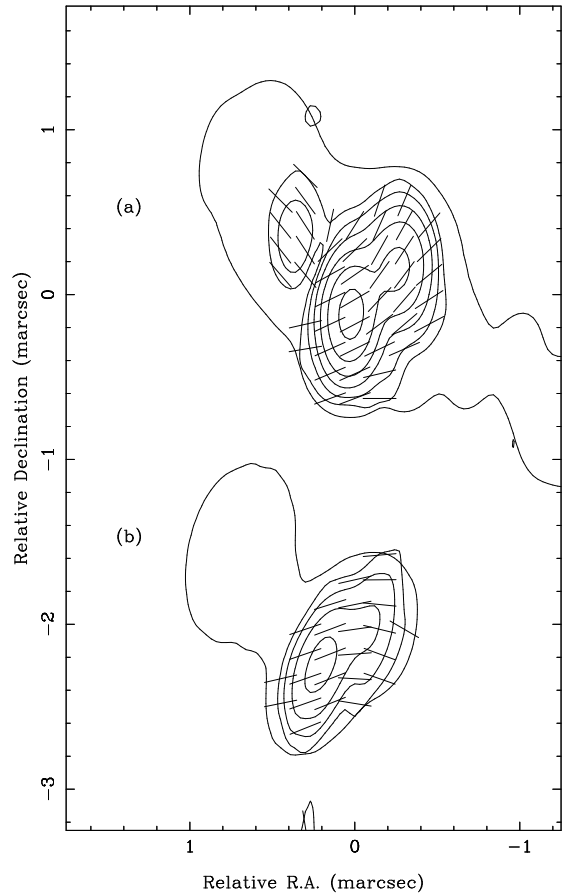}
\caption{Zoomed linear
polarization distributions from Figure~1 reveal more clearly the 
orientation of the EVPAs as discussed in \S 4; ($a$) is at 43~GHz and ($b$) is at
86~GHz, both showing every second contour.
\label{fig3}}
\end{figure}

\subsection{Total intensity images}
The 43 GHz images in Figure 1 show the well known core-jet structure of 3C\,273. The core is component D and the jet stretches to the south-west (and continues for another 20,000 mas in all wavebands). At this resolution, the jet is quickly resolved. At 86~GHz, the resolution is similar (because of the smaller array) but the lower sensitivity means that essentially only the first milliarcsecond of the jet can be seen. 

The component we label Q1/W1 is the same as component U4/K4 in Ojha et al (2004), who measured a proper motion of $9.9 \pm 0.4$c. Notice that the inner jet no longer points directly towards this component. This is because the direction of the inner jet varies on fairly short timescales over a range of structural position angles (SPAs) between at least $-110^{\circ}$ and $-140^{\circ}$ (e.g. Krichbaum et al. 1990). Abraham \& Romero (1999) proposed that the variation in orientation of the inner jet was periodic with a period of about 20 years, and they fitted a precessing jet model to the extant VLBI observations. It is now clear (Ojha et al 2003; Chen et al 2005) that the jet orientation varies on a timescale of months rather than decades, and is not periodic.

The bright inner jet components in Figure 1, components 6 and 7, are almost equidistant from the core at $r \sim 0.9$ mas, but at distinctly different structural position angles (SPAs). The northern component (6) is somewhat better aligned with the downstream jet components, and has a SPA of $-123^{\circ}$. The southern component (7) is stronger, has a flatter spectrum, and a SPA of $-144^{\circ}$. It is interesting that both are present simultaneously. We are not aware of earlier images that show such a morphology, and we conjecture that we have caught 3C\,273 "in the act" of changing its jet direction. If this is so, component 7, because of its location and spectrum, is presumably the younger component, and is defining a new jet direction. Since it has already drawn level with component 6, we expect component 7 to be moving faster. Proper motion measurements will easily support or refute these conjectures.

\subsection{Polarization images}
The polarized images shown in Figures~1 and 2 show that the inner jet is strongly polarized. Components Q6/W6 and Q7/W7 both exhibit fractional polarizations near $15\%$ at both frequencies. Two years earlier, Attridge (2001) found a component at a similar location, whose fractional polarization was $11\%$ at 86~GHz.
By contrast, linear polarization
is not detected in the core (component $D$) at either frequency. The limits
are $0.2\%$ at 43~GHz, and $1\%$ at 86~GHz. This also agrees with Attridge's 
(2001) observations. 

At lower frequencies, 3C\,273 also displays  moderate linear polarization in the jet, and little or no polarization in the core. These are properties common to many AGN (e.g.~Taylor 1998; Lister \& Smith 2000; Pollack et al. 2003; Ojha et al. 2004).

\section{Differential Faraday Rotation Across the Jet}
Although we are unable to determine the zero point for the EVPAs at 86 
GHz, Figure~3 shows clearly that there is a large {\em difference} 
in Faraday rotation between jet components 6 and 7. At 43~GHz the 
difference in EVPAs is
$\chi_{7}-\chi_{6}=+32.7^{\circ}$. At 86 GHz the 
difference in EVPAs is
$\chi_{7}-\chi_{6}=-33.9^{\circ}$. The difference between 
these numbers is $66.6^{\circ}=1.16$ radians, and this is independent of the 
zero point at 86~GHz. There is therefore an absolute {\em difference} in 
rotation measure between components 6 and 7 of $\sim3.2 \times 
10^{4}$~rad m$^{-2}$. In the source frame we multiply by $(1+z)^2$, which gives $\sim4.3 \times 10^{4}$~rad m$^{-2}$. 

The smallest possible rotation measures consistent with these data are
$-2.1\times 10^{4}$~rad m$^{-2}$ for component 7 and  $+2.1\times 10^{4}$~rad m$^{-2}$ for component 6. These are among the largest rotation measures ever observed. One or both of them are actually {\em lower} limits, because we have only determined their difference. The distance between components 6 and 7 is 0.3 mas. If we express the rotation measure difference as an angular gradient, then it is $1.3 \times 10^{5}$~rad m$^{-2}$ mas$^{-1}$.

Both Asada et al (2002) and recently Zavala \& Taylor (2005) have also measured differences in rotation measure across the jet of 3C\,273 at the same location, about 7 mas from the core, but at different epochs (1995.9 and two epochs in 2000 respectively). Asada et al found a gradient of about $80$~rad m$^{-2}$ mas$^{-1}$. Zavala \& Taylor find a gradient of $500$~rad m$^{-2}$ mas$^{-1}$, and suggest that Asada et al measured a smaller value because of lower resolution (their observations were made at 5 and 8~GHz). The transverse gradient at 7 mas from the core has the same sign as we find at 0.9 mas (more positive rotation measure on the north side of the jet), but it is smaller by 2-3 orders of magnitude. (We would require simultaneous observations to make the previous statement more precise.) 

It is interesting to compare the rotation measures found in the middle of each slice where the signal to noise ratio is highest and the effect of a gradient is negligible (so long as there is little depolarization). Asada et al measure about $350$~rad m$^{-2}$ at the center of their slice in 1995. This is consistent with the observations at 15 and 22~GHz throughout the following year reported by Ojha et al (2004). Zavala and Taylor measure about $800$~rad m$^{-2}$ at the same location some 4 years later. We suggest that all these measurements are in fact correct, and that the rotation measure distribution varies in time. This would also be consistent with the results of deep multiwavelength observations made by T. Chen (private communication), and with observations of the core region (see next section).

The Faraday rotation we observe must be external to components 6 and 7. Internal Faraday rotation will inevitably depolarize the source (sometimes called "front-back" depolarization) and there is no sign of this (see Table~1). In general the fractional polarization drops by a factor of two when the observed rotation reaches $\sim45^{\circ}$  (Burn, 1966). Here, the least depolarization would be observed if the rotation measures of components 6 and 7 were $\pm 2.1 \times 10^{4}$~rad m$^{-2}$. Even then, the predicted fractional polarizations at 43~GHz would be only $\sim 10\%$, whereas the observed values are close to $15\%$.
This rules out internal Faraday rotation by material that fills the jet, but Faraday rotation in a sheath or boundary layer is still a possibility. Inoue et al. (2003) reached the same conclusion for the gradient observed by Asada et al., and the same argument can be applied to Taylor \& Zavala's observations.

\section{Faraday Depolarization in the Core}
No linear polarization is detected from the core component D in these observations. The limits
are $0.2\%$ at 43~GHz, and $1\%$ at 86~GHz. 
The high rotation measures observed less than 1 milliarcsecond downstream from component D suggest that it is depolarized by differential Faraday rotation. This could be internal to the jet, but we favor an external screen because it is a natural extension of the screen that causes the Faraday rotation in front of components 6 and 7. If the intrinsic polarization of component D is $m_0$, and the variance in rotation measure along different lines of sight to component D is $\sigma^{2}_{RM}$, the observed polarization is given by $m(\lambda) = m_0 \exp(-2\sigma^{2}_{RM}\lambda^4)$ \citep{bur66}. We do not know the value of $m_0$, but a plausible range is from $0.7\%$ to $7.0\%$, since this is the range of polarizations observed by Nartallo et al (1998) at 270 GHz using the JCMT. The corresponding minimum values for $\sigma_{RM}$ are $1.6 - 8.0\times 10^4$ rad m$^{-2}$. (The first value is derived from the 43 GHz measurement.) We cannot be certain that the 270 GHz measurements by Nartallo et al (1998) actually refer to our component D. But if they do, we can derive an upper limit on $\sigma_{RM}$ from the requirement that the core is significantly polarized at 270~GHz: $\sigma_{RM} < 10^{6}$ rad m$^{-2}$. 

Note that this dispersion requires that the scale length for rotation measure fluctuations is smaller than component D itself. Farther out, at 0.9 mas, the fluctuation scale length is larger than the size of components 6 and 7 (or they would show depolarization at 43~GHz) but smaller than the distance between them. Similarly, at 7 mas the fluctuation scale length is larger than the restoring beam but smaller than the width of the jet.

Zavala and Taylor (2001) have also observed large and spatially resolved Faraday rotation in the core region of 3C\,273, which they found to be time-variable . At epoch 2000.07, they found a rotation measure of $-2.6 \times 10^{3}$~rad m$^{-2}$ in the north-east end of this region, near our component D, and $+1.9 \times 10^{3}$~rad m$^{-2}$ in the south-east end, near the location of our components 6 and 7. Their measurements (which were at 15 and 43~GHz) do not match ours in any obvious way, probably attesting to the variability of this source.

We note that if 3C\,273 were at a more typical redshift of, say, $z \simeq 1$, then components 6 and 7 would appear blended with the core, and could not be separated by current VLBI arrays. Then the apparent "core" would be polarized, and it would be very difficult to measure rotation measures as large as those reported here. Thus 3C\,273 may not be unusual in these properties; it is merely closer.

\section{Conclusions}
The results presented here demonstrate the value of polarization sensitive VLBI at 86~GHz. In particular, it allows us to probe the jet and its environment even closer to the central engine, and on the smallest available linear scales. It can also reveal very high rotation measures that cannot be seen at longer wavelengths because of source or bandwidth depolarization.

At 86~GHz, 3C\,273 reveals two strongly polarized ($\sim 16\%$) components within a milliarcsecond of the core that have rotation measures that differ by $\sim4.3 \times 10^{4}$~rad m$^{-2}$. This gradient is transverse to the jet direction and it has the same sign but is 2-3 three orders of magnitude larger than the gradients found by Asada et al (2000) and by Zavala \& Taylor (2005) at greater distances from the core. Extrapolating such a gradient to the core component readily accounts for its persistent lack of polarization even at 86 GHz.

The lack of depolarization in the jet requires that the Faraday rotation occurs in a region external to the visble jet itself, but presumably close to it. The data suggest that this Faraday rotating region is variable in time as well as spatially, and may be a boundary layer that represents the interaction of the jet with its environment. Rotation measure observations at the highest frequencies therefore offer a new way of exploring this environment.

\section{acknowledgments}

We thank Tingdong Chen for communicating the results of his observations. Radio astronomy at Haystack Observatory is supported by the NSF through the
Northeast Radio Observatory Corporation under grant AST-0096454. 
JFCW has been supported by NSF grant AST-00-98608.



\begin{thebibliography}{}
\bibitem[Abraham \& Romero(1999)]{ar99} Abraham, Z. \& Romero, G. E. 1999, \aap, 344, 61
\bibitem[Asada et al.(2002)]{asa02} Asada, K., Inoue, M., Uchida, Y., Kameno, S., Fujisawa, K., Iguchi, S. \& Mutoh, M. 2002, \pasj, 54, L39
\bibitem[Attridge(2001)]{att01} Attridge, J.~M. 2001, \apjl, 553, L31
\bibitem[Attridge, Greve, \& Krichbaum(1999b)]{att99b} Attridge, J.~M.,
     Greve, A., \& Krichbaum, T.~P. 1999b, in 2$^{nd}$ Millimeter-VLBI
     Science Workshop, ed. A.~Greve \& T.~P.~Krichbaum (Grenoble: IRAM), 13
\bibitem[Burn(1966)]{bur66} Burn, B.~J. 1966, \mnras, 133, 67
\bibitem[Cotton(1993)]{cot93} Cotton, W.~D. 1993, \aj, 106, 1241
\bibitem[Homan et al.(2001)]{hom01} Homan, D.~C., Ojha, R., Wardle,
     J.~F.~C., Roberts, D.~H., Aller, M.~F., Aller, H.~D., \& Hughes,
     P.~A. 2001, \apj, 549, 840
\bibitem[Homan et al.(2002)]{hom02} Homan, D.~C., Ojha, R., Wardle,
     J.~F.~C., Roberts, D.~H., Aller, M.~F., Aller, H.~D., \& Hughes,
     P.~A. 2002, \apj, 568, 99
\bibitem[Inoue et al.(2003)]{in03} Inoue, M., Asada, K. and Uchida, Y. 2003, in Radio Astronomy at the Fringe, eds. Zensus, J.~A., Cohen, M. H.  \& Ross, E. (ASP Conference Series, Vol. 300, Astronomical Society of the Pacific, San Francisco)
\bibitem[Krichbaum et al.(1990)]{kr90} Kritchbaum, T. P. \& 23 co-authors, 1990, \aap, 237, 3
\bibitem[Lister \& Smith(2000)]{lis00} Lister, M.~L., \& Smith, P.~S.
      2000, \apj, 541, 66
\bibitem[Nartallo et al.(1998)]{nar98} Nartallo, R., Gear, W. K., Murray, A. G.,       Robson, E. I. \& Hough, J. H. 1998, \mnras, 297, 667
\bibitem[Ojha et al.(2004)]{ojh04} Ojha, R., Homan, D.~C., Roberts, D.~H.,
Wardle, J.~F.~C., Aller, M.~F., Aller, H.~D., \& Hughes, P.~A. 2004,\apjs, 150, 187
\bibitem[Pollack et al.(2003)]{po03} Pollack, L. K., Taylor, G. B. \& Zavala, R. T. 2003, \apj, 589, 733
\bibitem[Roberts et al.(1994)]{rob94} Roberts, D.~H., Wardle,
      J.~F.~C., \& Brown, L.~F. 1994, \apj, 427, 718
\bibitem[Taylor(1998)]{tay98} Taylor, G.~B. 1998, \apj, 506, 637
\bibitem[Taylor \& Myers(2000)]{tay00b} Taylor, G.~B., \& Myers, S.~T.
2000, ``Polarization Angle Calibration Using the VLA Monitoring Program,''
VLBA Memo \#26
\bibitem[Zavala \& Taylor(2001)]{zt01} Zavala, R.~T. \& Taylor, G.~B. 2001, \apj, 550, L147
\bibitem[Zavala \& Taylor(2003)]{zt03} Zavala, R.~T. \& Taylor, G.~B. 2003, \apj, 589, 126
\bibitem[Zavala \& Taylor(2004)]{zt04} Zavala, R.~T. \& Taylor, G.~B. 2004, \apj, 612, 749
\bibitem[Zavala \& Taylor(2005)]{zt05} Zavala, R.~T. \& Taylor, G.~B. 2005, \apj, accepted, astro-ph/0505357

\end{thebibliography}
\end{document}